\documentstyle[aps,epsf,12pt]{revtex}
\begin{document}
\date{\null}
\title{Shot noise of Coulomb drag current
}
\author{V. L. Gurevich and M. I. Muradov}
\address{Solid State Physics Department, A. F. Ioffe Institute,
194021 Saint Petersburg, Russia}
\date{\today}
\maketitle
\begin{abstract}
We work out a theory  of shot noise in a special case. This is a
noise of the Coulomb drag current excited under the ballistic
transport regime in a one-dimensional nanowire by a ballistic
non-Ohmic current in a nearby parallel nanowire. We predict
sharp oscillation of the noise power as a function of gate
voltage or the chemical potential of electrons.  We also study
dependence of the noise on the voltage $V$ across the driving
wire. For relatively large values of $V$ the spectral density of
the low-frequency noise is proportional to $V^2$.
\end{abstract}

\section{Introduction}
The purpose of the present paper is to study the shot noise of the
Coulomb drag current in the course of ballistic (collisionless)
electron transport in a quantum nanowire due to a ballistic driving
current in an adjacent nanowire. The possibility of the Coulomb
drag effect in the ballistic regime has been demonstrated by
Gurevich, Pevzner and Fenton~\cite{GPF} for an Ohmic regime and by
Gurevich and Muradov~\cite{GM} for a non-Ohmic one and has been
experimentally observed by Debray {\it et al.}~\cite{DVR}.

If two wires, 1 and 2, are close and are parallel, the drag
force due to the ballistic current in wire 2 acts as a sort of
permanent acceleration on the electrons of wire 1 via the
Coulomb interaction. As a result, there appears a current noise
in wire 1 which depends on the voltage $V$ across wire 2. Such a
voltage-dependent noise can be looked upon as a sort of shot
noise. It is a theory of this noise that we will consider in the
present paper.

Let the two wires be much shorter than the electron mean free
path (typically a few $\mu$m).  Such nanoscale systems are
characterized by low electron densities, which may be varied by
means of the gate voltage. As in Ref.~\cite{GPF}, we assume the
wires to be of different widths though having the same lengths
$L$. Like in Refs.~\cite{GPF,GM} we consider the interaction
processes when electrons in nanowires 1 and 2 after scattering
remain within the initial subbands having the dispersion laws
\begin{equation}\label{0}
\varepsilon^{(1)}_{np}=\varepsilon^{(1)}_n(0)+p^2/2m,\quad
\varepsilon^{(2)}_{n^{\prime}p}=\varepsilon^{(2)}_{n^{\prime}}(0)+p^2/2m,
\end{equation}
where $p$ is the $x$-component of the electron quasimomentum,
the $x$ axis is parallel to the wires while $n$ and $n'$ are the
subbands' numbers.

The Boltzmann equation for fluctuations of the distribution
function describing electrons of wire 1 is
\begin{equation}\label{1}
\left({\partial\over{\partial
t}}+v{\partial \over{\partial x}}\right)\delta
F_{np}^{(1)}=-J_{p}\delta F_{np}^{(1)}+y_{np}^{(1)},
\end{equation}
where $\delta F^{(1)}$ are the fluctuations of the electron
distribution function in wire 1, and $y_{np}^{(1)}$ is the
Langevin random force originating in the interwire
electron-electron scattering. It is only such a scattering that we
will take into account in this otherwise ballistic system.

As in Refs.~\cite{GPF,GM}, we will solve Eq.~(\ref{1}) by
iterations. The collision term $J_p$ has the following
form
\begin{eqnarray}\label{sc_int}
&J_{p}&\Psi_{np}\nonumber\\
&=&\Psi_{np}\sum_{n'p'q}
W_{1pn,2p'n^{\prime}}^{1p+qn,2p'-qn^{\prime}}
\left[F^{(2)}_{n^{\prime}p'}
\left(1-F^{(1)}_{np+q}\right)\left(1-F^{(2)}_{n^{\prime}p'-q}\right)
+
\left(1-F^{(2)}_{n'p'}\right)F^{(1)}_{np+q}F^{(2)}_{n^{\prime}p'-q}\right]
\nonumber\\
&-&\sum_{n'p'q}\Psi_{np+q}
W_{1pn,2p'n^{\prime}}^{1p+qn,2p'-qn^{\prime}}
\left[F^{(2)}_{n^{\prime}p'-q}
\left(1-F^{(1)}_{np}\right)\left(1-F^{(2)}_{n^{\prime}p'}\right)
+
\left(1-F^{(2)}_{n'p'-q}\right)F^{(1)}_{np}F^{(2)}_{n^{\prime}p'}\right].
\end{eqnarray}
Here and henceforth it is implied that summation over $p$ or
$p'$ includes also the spin summation.

Applying the method of iterations we insert into the right-hand
side of Eq.~(\ref{1}) $\delta F_{np}^{(1)}$ in the zeroth
approximation given by the second and the third terms on the
right-hand side of Eq.~(\ref{6}) --- see below. We do not take
into account $\delta F_{np}^{(2)}$ as in this approximation the
functions $\delta F_{np}^{(1)}$ and $\delta F_{np}^{(2)}$
are uncorrelated being emitted by different reservoirs.

The correlation function of the Langevin forces can be obtained
using the procedure described in~\cite{GG,KS,MUR}
\begin{eqnarray}
\label{2}
\langle y_{np}^{(1)}({\bf r})y_{np'}^{(1)}({\bf
r}')\rangle_{\omega}=\delta_{{\bf r}{\bf
r}'}\sum_{n'p_1q}\left(\delta_{pp'}-\delta_{q,p'-p}\right)
W_{1pn,2p_1n^{\prime}}^{1p+qn,2p_1-qn^{\prime}}\nonumber\\
\times\left[F^{(1)}_{np}F^{(2)}_{n^{\prime}p_1}
\left(1-F^{(1)}_{np+q}\right)\left(1-F^{(2)}_{n^{\prime}p_1-q}
\right)\right.\left.+F^{(1)}_{np+q}F^{(2)}_{n^{\prime}p_1-q}
\left(1-F^{(1)}_{np}\right)\left(1-F^{(2)}_{n^{\prime}p_1}\right)\right].
\end{eqnarray}
Here the first term on the right-hand side, including
$\delta_{pp'}$, is a sum of the `in' and `out' terms in the
Boltzmann equation for the state $np$, whereas the other term is the
sum of collision probabilities where one of the initial states
is $np$ and one of the final states is $np'$ {\it et vice
versa.} The correlation function, as well as the scattering
integral, satisfy the following relations representing the
particle number conservation
\begin{equation}\label{3}
\sum_{np}\langle y_{np}^{(1)}({\bf r})y_{np'}^{(1)}({\bf
r}')\rangle_{\omega}= \sum_{np'}\langle y_{np}^{(1)}({\bf
r})y_{np'}^{(1)}({\bf r}')\rangle_{\omega}=0,\quad
\sum_pJ_p\Psi_{np}=0.
\end{equation}

The solution of Eq.(\ref{1}) can be written using the Green
function satisfying the equation
\begin{equation}\label{4}
\left({\partial\over{\partial t}}+v{\partial \over{\partial
x}}\right)G(xt|x't')=\delta(x-x')\delta(t-t').
\end{equation}
We have
\begin{equation}\label{5}
G(xt|x't')={1\over{|v|}}\delta\left(t-t'-{x-x'\over{v}}\right)
\theta\left({x-x'\over{v}}\right),
\end{equation}
where $v=p/m$ is the $x$-component of the electron velocity,
so that
\begin{eqnarray}
\delta
F_{np}(x,t)=\int_{-\infty}^{+\infty}dt'
dx'G(xt|x't')\left\{-J_{p}\delta F_{np}+y_{np}(x',t')\right\}\nonumber\\
+|v|\int_{-\infty}^{+\infty}dt'G\left(xt\left|-{L\over2},t'\right.
\right)\delta
F_{np}\left(-{L\over2},t'\right)+
|v|\int_{-\infty}^{+\infty}dt'G\left(xt\left|{L\over2},t'\right.
\right)\delta
F_{np}\left({L\over2},t'\right).
\label{6}
\end{eqnarray}
For the current fluctuations we get
\begin{equation}\label{7}
\delta I(x,t)={e\over{L}}\sum_{np}v\delta
F_{np}(x,t)
\end{equation}
 The
collision integral takes into account only the interwire
electron-electron scattering.

Using the particle number conserving property~(\ref{3})
we write
\begin{eqnarray}
\delta I(x,t)&=&{e\over
L}\sum_{np>0}
\int_{-L/2}^{L/2}dx'\left\{-J_{p}\delta F_{np}\left(x',t-
{x-x'\over{v}}\right)+
y_{np}\left(x',t-
{x-x'\over{v}}\right)\right\}\nonumber\\
&+&{e\over L}\sum_{np>0}v\delta
F_{np}\left(-{L\over2},t-{x+L/2\over{v}}\right)
+{e\over L}\sum_{np<0}v\delta
F_{np}\left({L\over2},t-{x-L/2\over{v}}\right).
\label{9}
\end{eqnarray}
Here the first term on the right-hand side describes the noise
induced by the collisions of the electrons in wire 1 with the
electrons in wire 2 while the rest two terms describe the
contributions of fluctuations of distributions for the electrons
entering wire 1 from the left and the right boundaries.
Accordingly, the noise power consists of the Nyquist noise and
the nonequilibrium noise due to collisions with the
nonequilibrium electrons in wire 2.

Using the identity
\begin{eqnarray}
\langle\delta F_{np}\left(x,t\right)\delta
F_{np'}\left(x',t'\right)\rangle=
L\delta_{pp'}F_{np}\left(1-F_{np}\right)\delta\left[x-x'-v_p(t-t')\right]
\end{eqnarray}
we have for the total power of the low frequency noise $P_{\rm
tot}=P_{\rm N}+P$ where the power of the equilibrium Nyquist noise is
$$ P_{\rm N}={e^2\over{\pi\hbar}}\sum_n\int_0^{\infty}dpv\{f_L
(\varepsilon_{np}-\mu)
[1-f_L(\varepsilon_{np}-\mu)]
+f_R(\varepsilon_{np}-\mu)[1-f_R(\varepsilon_{np}-\mu)]\},
$$
or
\begin{equation}
P_{\rm N}=4GT,\;\;\;G={e^2/{\pi\hbar}}N.
\end{equation}
Here $N$ is the number of open channels in wire 2.
We will write throughout the paper $T$ instead of $k_{\rm B}T$
where $T$ is the temperature. The noise power due
to the drag is
\begin{equation}\label{10_1}
P=P_S+P_L+P_R,
\end{equation}
where
\begin{equation}\label{10_2}
P_S=2e^2\sum_{n,p>0,p'>0}<y_{np}y_{np'}>_{\omega}
\end{equation}
describes
fluctuations induced by the sources in the wire,
\begin{equation}\label{10_3}
P_L=-
4e^2\sum_{n,p'>0}\sum_{p>0}
F_{np}^{(1)}\left(1-F_{np}^{(1)}\right)
J_{p'}\delta_{pp'}
\end{equation}
describes the scattering of fluctuations entering
the wire from the left reservoir, and
\begin{equation}\label{10_4}
P_R=
4e^2\sum_{n,p'>0}\sum_{p<0}
F_{np}^{(1)}\left(1-F_{np}^{(1)}\right)
J_{p'}\delta_{pp'}
\end{equation}
describes the scattering of fluctuations entering
the wire from the right reservoir.

The scattering probability is
\begin{equation}\label{sp}
W_{1pn,2p^{\prime}n^{\prime}}^{1p+qn,2p^{\prime}-qn^{\prime}}\nonumber\\
={2\pi\over{\hbar}}
\left|V_{1pn,2p^{\prime}n^{\prime}}^{1p+qn,2p^{\prime}-qn^{\prime}}\right|^2
\delta(\varepsilon^{(1)}_{np}+\varepsilon^{(2)}_{n^{\prime}p^{\prime}}-
\varepsilon^{(1)}_{np+q}-\varepsilon^{(2)}_{n^{\prime}p^{\prime}-q}).
\end{equation}
It can be transformed with the help of relations (see Ref.~\cite{GPF})
\begin{equation}\label{delta_func}
\delta(\varepsilon^{(1)}_{np}+\varepsilon^{(2)}_{n^{\prime}p^{\prime}}-
\varepsilon^{(1)}_{np+q}-\varepsilon^{(2)}_{n^{\prime}p^{\prime}-q})=
{m\over{|p-p^{\prime}|}}\delta(q+p-p^{\prime})
\end{equation}
and
\begin{equation}\label{matr_el}
\left|\langle
np-q,n'p|V|np,n'p-q\rangle\right|^2=\left({2e^2\over{\kappa
L}}\right)^2 g_{nn^{\prime}}(q) \end{equation} where
\begin{equation}\label{gnn} g_{nn^{\prime}}(q)=\left(\int d{\bf
r}_{\perp}\int d{\bf r}^{\prime}_{\perp} |\phi_n({\bf r}_{\perp})|^2
K_0\left({|q|{\hbar}^{-1}}|{\bf r}_{\perp}-{\bf
r}^{\prime}_{\perp}|\right) |\phi_{n^{\prime}}({\bf
r}^{\prime}_{\perp})|^2\right)^2,
\end{equation}
$\phi_n({\bf r}_{\perp})$ being the functions of transverse
quantization. We assume, in the spirit of the
Landauer-B\"uttiker-Imry~\cite{LB} approach, wire 1 to be
connected to reservoirs which we call `left' $(l)$ and `right'
$(r)$, each of these being in independent equilibrium described
by the shifted chemical potentials $\mu^{(l)}=\mu-\Delta\mu/2$
and $\mu^{(r)}=\mu+\Delta\mu/2$.  Here $\mu$ is the average
chemical potential while $\Delta\mu/e=V$ is the voltage across
wire 2 (we will assume that $eV>0$) and $e<0$ is the electron
charge. Therefore, the electrons entering the wire from the
`left' (`right') and having quasimomenta $p^{\prime}>0$
[$p^{\prime}<0$] are described by
$F^{(2)}_{n^{\prime}p^{\prime}}=
f(\varepsilon^{(2)}_{n^{\prime}p^{\prime}}-\mu^{(l)})$
$\left[F^{(2)}_ {n^{\prime}p^{\prime}}=
f(\varepsilon^{(2)}_{n^{\prime}p^{\prime}}-\mu^{(r)})\right]$
respectively. For the noise power $P_S$ we get
\begin{equation}\label{np}
P_S=2e^2m{2\pi\over{\hbar}}\left({L\over{2\pi\hbar}}\right)
\left({2e^2\over{\kappa
L}}\right)^2\left({2L\over{2\pi\hbar}}\right)^2
\sum_{nn'}\int_0^{\infty}dp\int_0^{\infty}dp'
{g_{nn^{\prime}}(p+p')\over{p+p'}}{\cal S}
\end{equation}
where
\begin{eqnarray}\label{20}
{\cal S}&=&f(\varepsilon_{np'}-\mu)\left[1-f\left(\varepsilon_{n'p'}-
\mu-{eV\over2}\right)\right]\\\nonumber
&\times&f\left(\varepsilon_{n'p}-\mu+{eV\over2}\right)
[1-f(\varepsilon_{np}-\mu)]
\left(1+\exp{eV\over T}\right)
\end{eqnarray}
One can take out of the integral all of the slowly varying
functions. Exploiting the relation
\begin{equation}\label{integral}
\int_{-\infty}^{+\infty}dx{\exp{(x+2a)}\over{\left[1+\exp{(x+2a)}\right]
\left[1+\exp{(x+2b)}\right]}}={(a-b)}{\exp{\left(\displaystyle{a-b}\right)}
\over{\sinh{\displaystyle{(a-b)}}}}
\end{equation}
one gets
\begin{eqnarray}\label{main_formula}
P_S=-2eJ\coth{\left({eV\over 2T}\right)}.
\end{eqnarray}
The drag current $J$ is according to Ref.~\cite{GM}
\begin{equation}\label{curr_formula}
J={1\over 2}J_0\sinh x^{(0)}\cdot
{{x^{(-)}_{nn'}}\over{\sinh{x^{(-)}_{nn'}}}}\cdot
{{x^{(+)}_{nn'}}\over{\sinh{x^{(+)}_{nn'}}}}.
\end{equation}
Here we have
introduced notation $x^{(0)}=eV/2T$, $x^{(\pm)}_{{nn'}}=eV/4T
\pm{\varepsilon_{nn'}/2T}$,
\begin{equation}\label{J_0}
J_0=-{8e^5m^3LT^2\over{\kappa^2\pi^2\hbar^4}}\cdot
{g_{nn'}(2p_n)\over{p_n^3}}
\end{equation}
(where $\kappa$ is the dielectric susceptibility) and
\begin{equation}\label{E_nn}
\varepsilon_{nn'}=\varepsilon_{n}^{(1)}(0)-\varepsilon_{n'}^{(2)}(0),\quad
mv_n=p_n=\sqrt{2m[\mu-\varepsilon_{n}^{(1)}(0)]}.
\end{equation}
According to Refs.~\cite{GPF,GM}, the current $J$ as a function
of the gate voltage comprises a system of spikes; the position
of each spike is determined by a coincidence of a pair of levels
of transverse quantization, $\varepsilon_{n}(0)$ and
$\varepsilon_{n'}(0)$ in both wires.  Using the explicit form of
the scattering operator Eq.(\ref{sc_int}) one can show that the
noise power $P_R$ can be expressed through $P_L$ simply by the
replacement $eV\rightarrow\,-eV$.  Indeed, taking into
consideration that summation in (\ref{10_3}) is performed over
the positive quasimomenta $p>0$ and using the expression for the
scattering probability (\ref{sp}), (\ref{delta_func}) we get
\begin{eqnarray}\label{summa_1}
\sum_{p'>0}J_{p'}\delta_{pp'}=\left({2e^2\over{\kappa
L}}\right)^2m
{L\over{2\pi\hbar}}\sum_{n',p'<0}{g_{nn'}(p'-p)\over{|p-p'|}}
{\cal F},
\end{eqnarray}
\begin{equation}
{\cal F}=\left[F^{(2)}_{n^{\prime}p'}
\left(1-F^{(1)}_{np'}\right)\left(1-F^{(2)}_{n^{\prime}p}\right)+
\left(1-F^{(2)}_{n'p'}\right)F^{(1)}_{np'}F^{(2)}_{n^{\prime}p}\right]
\end{equation}
in the expression for $P_L$ while in the expression for $P_R$
(taking into account that there the sum is over the negative
$p$) we have
\begin{eqnarray}\label{summa_2}
\sum_{p'>0}J_{p'}\delta_{pp'}=-\left({2e^2\over{\kappa
L}}\right)^2m
{L\over{2\pi\hbar}}\sum_{n',p'>0}{g_{nn'}(p'-p)\over{|p-p'|}}
{\cal F}
\end{eqnarray}
Now replacing $F^{(2)}_{sk}$ by
$f(\varepsilon_{sk}-\mu+eV/2)$ and
$f(\varepsilon_{sk}-\mu-eV/2)$ for $p>0$ and $p<0$
respectively, for the $P_L$ we
have
\begin{equation}\label{np_l}
P_L=-4e^2m{2\pi\over{\hbar}}\left({L\over{2\pi\hbar}}\right)
\left({2e^2\over{\kappa
L}}\right)^2\left({2L\over{2\pi\hbar}}\right)^2
\sum_{nn'}\int_0^{\infty}dp\int_0^{\infty}dp'
{g_{nn^{\prime}}(p+p')\over{p+p'}}{\cal L}
\end{equation}
where
\begin{eqnarray}\label{20_l}
{\cal L}&=&f\left(\varepsilon_{np}-\mu\right)
\left[1-f\left(\varepsilon_{np}-\mu\right)\right]\\\nonumber
&\times&\left\{
f\left(\varepsilon_{n'p}-\mu+{eV\over2}\right)
f\left(\varepsilon_{np'}-\mu\right)
\left[1-f\left(\varepsilon_{n'p'}-\mu-{eV\over2}\right)\right]
\right.\\\nonumber
&+&\left.f\left(\varepsilon_{n'p'}-\mu-{eV\over2}\right)
\left[1-f\left(\varepsilon_{np'}-\mu\right)\right]
\left[1-f\left(\varepsilon_{n'p}-\mu+{eV\over2}\right)\right]\right\}
\end{eqnarray}
Finally we get
\begin{equation}\label{p_left}
P_L=2eJ\left[
{1\over x_{nn'}^{(-)}}-
{1\over\sinh x^{(0)}}\cdot
{{\sinh x_{nn'}^{(+)}
\over{\sinh x_{nn'}^{(-)}}}}\right]
\end{equation}
The sum of $P_L$ and $P_R$ is
\begin{equation}\label{p_l_r}
P_L+P_R=-2eJ\left\{{1\over{\sinh{x^{(0)}}}}
\left[{\sinh{x_{nn'}^{(+)}}\over{\sinh{x_{nn'}^{(-)}}}}+
{\sinh{x_{nn'}^{(-)}}\over{\sinh{{x_{nn'}^{(+)}}}}}\right]
-{x^{(0)}\over {x_{nn'}^{(-)}x_{nn'}^{(+)}}}\right\}
\end{equation}
where $J$ is given by Eq.~(\ref{curr_formula}).
For $eV\ll T$ one gets for the variation of the total noise
power in wire 1 due to the presence of wire 2
\begin{equation}\label{11}
P=P_S+P_L+P_R=eJ_0 {\left[\displaystyle
     {\varepsilon_{nn'}\over
2T}\right]^2\cdot{\left[\sinh\left({\displaystyle
      {\varepsilon_{nn'}\over 2T}}\right)\right]^{-2}}}.
\end{equation}
One can verify that this result is consistent with the
fluctuation-dissipation theorem. Indeed, the linear response to
the voltage $V^{(1)}$ across wire 1 in this case can be written as
\begin{equation}\label{fd1}
J=\left(G-G_{\rm tr}\right)V^{(1)}.
\end{equation}
Here, according to Eq.(\ref{curr_formula}),
we have introduced the transconductance $G_{tr}$
\begin{equation}\label{fd2}
G_{\rm tr}=-{e\over{4T}}J_0 {\left[\displaystyle
     {\varepsilon_{nn'}\over
2T}\right]^2{\left[\sinh\left({\displaystyle
      {\varepsilon_{nn'}\over 2T}}\right)\right]^{-2}}}.
\end{equation}
We assume that $J$, $V^{(1)}$ (and, of course, $G_{\rm tr}$) are
positive quantities; then the fluctuation-dissipation theorem states
that the additional equilibrium contribution to the noise due to the
wire 2 is
\begin{equation}\label{fd3}
P=-4G_{\rm tr}T,
\end{equation}
which coincides with Eq. (\ref{11}).

Let us consider in detail the opposite case $eV\gg T$.
In this case one gets a nonvanishing result for
Eq.(\ref{curr_formula}) only if $|\varepsilon_{nn\prime}|\,<\,eV/2$
and one obtains the Poisson limit for the noise power
\begin{equation}\label{large_1}
P=-2eJ,
\end{equation}
where the drag current is given by
\begin{equation}\label{large_2}
J={\cal B}
\left[\left({eV\over2}\right)^2-
\left({\varepsilon_{nn'}}\right)^2\right],\quad
{\cal B}=-{2e^5m^3L\over{\kappa^2\pi^2\hbar^4}}
\cdot\displaystyle{g_{nn^{\prime}}(2p_n)\over{p_n^3}}.
\label{12}
\end{equation}

The situation is illustrated in Figure~1, where we plot the noise
power versus applied driving voltage for various values of the
difference of transverse energy levels $\varepsilon_{nn'}$. For
the small values of $eV/T$ we have an additional contribution to
the thermal (equilibrium) noise due to wire 2 described by Eq.
(\ref{11}), while for the large values of $eV/T$ the
contribution to the (now nonequilibrium) noise in the drag wire
is described by a quadratic dependence of Eq.(\ref{large_2}).
To clear the situation further we plot the dependence of the
noise power on the difference of transverse energy levels
$\varepsilon_{nn'}$ in Figure~2.

In summary, we have developed a theory of a shot noise in a
quantum wire excited by a non-Ohmic current in a nearby parallel
nanowire. A ballistic transport in both nanowires is assumed.
The shot noise power $P$ as a function of the gate
voltage comprises a system of spikes; the position of each spike
is determined by a coincidence of a pair of levels of transverse
quantization, $\varepsilon^{(1)}_{n}(0)$ and
$\varepsilon^{(2)}_{n'}(0)$ in both wires. For $eV\gg T$, $P$
is a quadratic function of the driving voltage $V$. The effect
may play an important role in the investigation of the interwire
Coulomb scattering as well as 1D band structure of the wires.

The authors are grateful to P. Debray for sending them a
preprint of paper~\cite{DVR} prior to publication. The authors
are pleased to acknowledge the support for this work by the
Russian National Fund of Fundamental Research (Grant
\#~97-02-18286-a).

\newpage
\centerline{FIGURE CAPTIONS}
\bigskip

1. Fig. 1. Noise power $P$ in wire 1 as a function of the voltage
$V$ applied across wire 2 for
$\varepsilon_{nn'}/T=0,\,1,\,2,\,3$. As $\varepsilon_{nn'}/T$
goes up the curves are shifted to the right in the upper part of
the figure.

2. Fig. 2. Noise power $P$ as a function of
$\varepsilon_{nn'}/T$ for $eV/T=0,\,3,\,6,\,9,\,12$. The values
of the functions for $\varepsilon_{nn'}=0$ go up as $eV/T$ goes up.

\end{document}